\documentclass{emulateapj}

\usepackage{natbib}
\usepackage{graphicx}
\usepackage{amssymb}

\usepackage[dvipsnames]{color}

\shorttitle{DUST-DRIVEN WIND FROM DISK GALAXIES}
\shortauthors{Sharma, Nath, Shchekinov}

\begin{document}


\newcommand{\3}{\ss}
\newcommand{\n}{\noindent}
\newcommand{\eps}{\varepsilon}
\def\be{\begin{equation}}
\def\ee{\end{equation}}
\def\ba{\begin{eqnarray}}
\def\ea{\end{eqnarray}}
\def\de{\partial}
\def\msun{M_\odot}
\def\div{\nabla\cdot}
\def\grad{\nabla}
\def\rot{\nabla\times}
\def\ltsima{$\; \buildrel < \over \sim \;$}
\def\simlt{\lower.5ex\hbox{\ltsima}}
\def\gtsima{$\; \buildrel > \over \sim \;$}
\def\simgt{\lower.5ex\hbox{\gtsima}}
\def\etal{{et al.\ }}
\def\red{\textcolor{red}} 
\def\blue{\textcolor{blue}}


\title{DUST-DRIVEN WIND FROM DISK GALAXIES}

\author{Mahavir Sharma$^1$, Biman B. Nath$^1$, Yuri Shchekinov$^2$}
\affil{$^1$ Raman Research Institute, Sadashiva Nagar, Bangalore 560080, India}
\affil{$^2$ Department of Physics, Southern Federal University,
        Rostov on Don, 344090 Russia}
\email{mahavir@rri.res.in; biman@rri.res.in; yus@sfedu.ru}

\begin{abstract}
We study gaseous outflows from disk galaxies driven by radiation pressure on 
dust grains. We include the effect of bulge and dark matter halo and show that
the existence of such an outflow implies a maximum value of disk mass-to-light
ratio. We show that the terminal wind speed is proportional to the disk rotation
speed in the limit of a cold gaseous outflow, and that in general there is a 
contribution from the gas sound speed. Using the mean opacity of dust grains and 
the evolution of the luminosity of a simple stellar population, we then show that 
the ratio of the wind terminal speed ($v_\infty$) to the galaxy rotation speed ($v_c$) 
ranges between $2 \hbox{--} 3$ for a period of $\sim 10$ Myr after a burst of star
formation, after which it rapidly decays. This result is independent of any
free parameter and depends only on the luminosity of the stellar population 
and on the relation between disk and dark matter
halo parameters. 
We briefly discuss the possible implications of our results.
\end{abstract}

\keywords{
galaxies: starburst  --- galaxies: evolution ---intergalactic medium
}

\section{Introduction}
Galactic outflows are believed to be an important ingredient
in the evolution of galaxies. It
is thought to provide a feedback
mechanism for the regulation of star formation in galaxies, and
the subsequent chemical evolution in them. Also,
galactic scale winds are thought to enrich the intergalactic medium (IGM)
with metals. Observations also show that galactic scale outflows are
common in starburst galaxies in both local and high redshift universe (Veilleux et al. 2005).

In the standard scenario, the interstellar
medium (ISM) of the starburst galaxy is heated by supernovae (SN),
the thermal pressure of the hot gas drives the outflowing gas. Larson (1974),
Saito (1979), Dekel \& Silk (1986) argued that supernovae in a star 
forming galaxy would drive an outflow in excess of
the escape speed from the galaxy.
This scenario, however, has met with problems from new observations. Recent
observations of cold clouds embedded in the hot gas show that
the maximum speed of these clouds is correlated with the star
formation rate (Martin 2005), a correlation that is more easily
explained by outflows driven by radiation pressure than thermal pressure.
Observations of Lyman break galaxies (LBG) at $z\sim 3$
have also found a correlation between the outflow speed and
the star formation rate, as well as with the reddening due to dust.
Also, SN explosions in disk galaxies may only produce a thickening
of the disk gas because of the incoherent nature
of these explosions (Fragile, Murray, Lin 2004).
Moreover, cold clouds embedded in the hot gas in the outflow are not
likely to survive longer than a Myr, because of 
various instabilities and/or evaporation by thermal
conduction. 

Murray \etal (2005; hereafter referred to as MQT05) 
showed that radiation pressure can be comparable
to the ram pressure in hot outflows, and considered
radiatively driven shells of gas and dust. Martin (2005) found that this
scenario provides a better framework in order to understand the observed
correlations. Nath \& Silk
(2008) discussed a hybrid model of outflows with radiation and thermal
pressure.

In this paper, we study dust driven gaseous winds from luminous disk
galaxies. We estimate the
terminal speed of the combined gas and dust flow, taking into account the
gravity of the dark matter halo, and the disk mass-to-light ratio that
is expected from stellar population synthesis models.

\section{Dust-driven wind from a Keplerian disk}
Recently Zhang \& Thompson (2010) considered a disk of radius $d$
constant surface density ($\Sigma$) and surface brightness ($I$).
Assuming a cylindrical geometry, the force
of gravitation $f_{g}(z)$, and that due to radiation 
$f_{r} (z)$, along the pole are given by,
\be
f_{g}=2 \pi G \Sigma\int\frac{zrdr}{(r^{2}+z^{2})^{3/2}} \,,
f_{r}=\frac{2 \pi\kappa I}{c}\int\frac{z^{2}rdr}{(r^{2}+z^{2})^{2}} \,,
\label{eq:fg}
\ee
where $\kappa$ is the average opacity of dust grains to absorption
and scattering of photons. 
The ratio of these forces,
the Eddington ratio, increases with the height $z$, beginning with a value
of $\Gamma_0 = {\kappa I \over 2 c G \Sigma}$ at the disk centre at $z=0$.
From energy conservation they found the velocity at height $z$, to be,

\be
v_{z}^{2}=4\pi G \Sigma d\left(\Gamma_0 \tan ^{-1}\left(\frac{z}{d}\right)-1-\frac{z}{d}+\sqrt{1+\frac{z^{2}}{d^{2}}}\right)\,.
\ee
This implies a terminal velocity of $v_\infty\approx \sqrt{4 \pi G \Sigma
d (\pi\Gamma_0 /2 -1)}$, with a lower limit on $\Gamma_0 \sim 2 /\pi$
for driving an outflow.

Assuming a Keplerian disk, with average rotation speed of
$\langle v_{c,kep} \rangle \sim
{1 \over 2} \sqrt{\pi G \Sigma d}$,
the terminal velocity can be written as $
v_{\infty}=4(\sqrt{\frac{\pi\Gamma_0}{2}-1}) \langle v_{c,kep} \rangle$,
which yields a value $v_{\infty}\simeq\ 3 \langle v_{c,kep} \rangle $, 
for $\Gamma_0=1$. 
But these results will change with the proper inclusion of bulge and dark matter halo.


\section{Gaseous outflows}

In this paper, we discuss steady, rotating wind in which dust is
propelled outwards by radiation
pressure and drags the gas with it. We ignore magnetic forces and treat the wind as
a single-phase fluid,
which is marginally optically thick (MQT05).

Although our results are valid for all temperatures, in practice the model allows
for only cold flows (as observed
in NaI and MgII lines (e.g. Martin 2005)) for the following reasons. The sputtering
radius for grains over
a wind timescale of 10 Myr (which we derive below) is $a \sim 50 \times \frac{n}{0.1/cc}$ $\AA$ for
$T \sim 10^5$ K (Tielens
\etal 1994), which implies that only grains smaller than $5$ $\AA$ are destroyed,
without changing the opacity. But the radiative cooling time is small ($\le 1$ Myr), so for adiabatic approximation to be valid, the adiabatic-cooling time scale should be shorter than the radiative-cooling timescale, which limits the temperature $\le 10^4$ K. Also the timescale for dust-heating is $\sim10$ Myr and can be neglected.



Secondly, we improve upon the previous estimates of the required
value of $\Gamma_0$ needed to drive a wind, by taking into account 
the gravity of bulge and halo. Clearly
the relation between $v_\infty$ of the wind and $v_{c}$ in which the 
rotation speed is estimated from a keplerian disk is not relevant for
disk galaxies, because the rotation speed must be calculated from the 
dark matter halo.

\subsection{Disk, bulge and halo parameters}
We consider a disk with constant surface density ($\Sigma$) and surface brightness ($I$), 
of radius $d$, and which is embedded in a bulge and a halo. 
We assume a spherical mass distribution in the bulge and
the halo. For the bulge, we assume a total mass of $M_b$ inside a radius
$r_b \ll d$. 

For the halo, we consider a Navarro-Frenk-White (NFW) profile, 
with total mass $M_h$ characterized
by a concentration parameter $c=r_{vir}/r_s$ (Navarro, Frenk, White 1997). 
We fix the total halo mass for a given disk
mass ($M_d=\pi d^2 \Sigma$), by the ratio $M_h/M_d\sim 1/0.05$, as determined by
Mo, Mao \& White (1998) (referred to as MMW98 hereafter). 
We use the prescription of MMW98 for the disk exponential scale-length
 $R_d \sim (1/\sqrt{2}) \lambda R_{200}
f_c^{-1/2} f_R$, using their eqns 23, 26, 32, and use 
$d=R_d/\sqrt{2}$, since the total
masses in the case of uniform density and 
exponential disk are given by $M=\pi d^2 \Sigma=2 \pi 
\Sigma_0 R_d^2$.
The rotation speed implied by the NFW profile peaks at a radius $r\sim 2 r_s$, given by,
\be
v_c ^2=v_{200}^2 \,{c \over 2} \, {\ln (3)-2/3 \over \ln (1+c) -c/(1+c)} \,,
\ee
where $r_s$ is the scale radius of NFW profile and $v_{200}$ is the rotation speed at the virial radius.
We choose this value of the maximum rotation speed
to represent the $v_c$ of the disk galaxy, since Figure 2 of MMW98 shows that the value of $v_c$
from the flat part of the total rotation curve does not differ much from the peak of the
rotation curve from halo only. 
We also use $r_b/d\sim 0.1$, and $M_b/M_d \sim 0.5$, 
consistent with observed  range
of luminosity ratio between bulge and
disk (Binney \& Merrifield 1998).

\subsection{Wind terminal speed}
To determine the terminal speed of the wind, we use the fact that the Bernoulli
function is preserved along a streamline, assuming that a streamline extends
from the base to infinity. In an isothermal wind the terminal speed tends to infinity
as the wind maintains constant sound speed. 
It is however more reasonable to assume
a polytropic equation of state. 

One can write the Bernoulli equation for a polytropic gas (with adiabatic index $\gamma$)
along a streamline: ${v^2 \over 2} + {c_s^2 \over \gamma -1} + \phi= E$, where $c_s$ is the sound
speed, $\phi$ is the potential and $E$ is a constant. Equating the values at the base and infinity,
we have
\be
\frac{v_\infty ^2}{2}+{c_{s,\infty} ^2\over \gamma -1}+\phi _\infty=\frac{v_b^2}{2} +{c_{s,b} ^2\over \gamma -1}  + \phi_b  \,,
\label{eq:bern}
\ee
where $v_b$ is the wind speed at the base ($z=0$), $c_{s,b} $ is the sound 
speed at the base, and 
$c_{s,\infty}$is the sound speed at infinity, which is negligible. The potential is,  

\begin{eqnarray}
\phi=&&-2\pi G \Sigma d \Gamma_0 \tan ^{-1} (z/d)+2\pi G \Sigma d \left(\frac{z}{d}-\sqrt{1+\frac{z^{2}}{d^{2}}}\right) \nonumber\\
&& +\frac{L^{2}}{2 r^{2}}-{G M_b \over \sqrt{r^2 +z^2}} -{G M_s \over r_s} {\ln (1+\frac{\sqrt{r^2 +z^2}}{r_s}) \over
\frac{\sqrt{r^2 +z^2}}{r_s}} \,.
\end{eqnarray}

Here the first term denotes a pseudo-potential due to radiation pressure. The second term 
refers to the gravitational potential 
of the disk, and here we have assumed for analytical simplicity
that gas stays near the pole, so these forces are given by eqn$(\ref{eq:fg})$. The third term is the potential due to the outward centrifugal force of the rotating gas in the wind, and the last two terms denote the effect of 
the bulge and halo gravity.
We also assume that the bulge exerts a negligible radiation pressure, since
the dominant bulge stellar population is old and red and the mean opacity $\kappa$ of dust grains
in these wavelengths is smaller than in blue band.

The centrifugal force at the base of the rotating
 wind should be equal to gravitational force
due to the mass inside $a$, the radial distance at the base. We
assume that $a \ge r_b$, the bulge radius, while
being much smaller than $d$, the disk radius. Hence the centrifugal force is written as,
\begin{equation}
 \frac{L_{base}^{2}}{a^{3}}=\frac{G M_{b}}{a^{2}}+\frac{G M_{s}}{a^{2}}\left[\ln(1+\frac{a}{r_s})-\frac{a}{a+r_s}\right]
\end{equation}
The first term on RHS is due to the bulge, and the second term, due to the dark matter halo. For $a \ll r_s$, the second term can be approximated as $\frac{G M_s}{2 r_s^{2}}$ which is much smaller than the bulge term $ \frac{G M_{b}}{a^{2}}$, because the gravity of baryons dominates at small radii.
Neglecting the halo term, the specific angular momentum is given by,
\begin{equation}
 L_{base}=\sqrt{GM_{b} a} \,.
\end{equation}
Finally, the values of the potential at the base and infinity are as follows,

\begin{eqnarray}
\phi_b=&&-2 \pi G \Sigma d +{G M_b \over 2 a}-{G M_b \over a} -{G M_s \over r_{s}}\left[\frac{\ln (1 + a /r_s)}{\frac{a}{r_{s}}}\right] \nonumber\\
\phi_\infty=&&-2 \pi G \Sigma d \Bigl ({\Gamma_0 \pi \over 2} \Bigr ) \,.
\end{eqnarray}
We note that
the radiation pseudo-potential is zero at the base but $\sim -2 \pi G \Sigma d ({\Gamma_0 \pi \over 2})$ at infinity along the pole. Since the dark matter
halo is truncated at $r_{200}$, its potential vanishes at infinity.

It is reasonable to assume that the wind speed at the base is comparable to
the sound speed in the disk ($v_b \sim c_{s,b}$). 
Putting it all in eqn$(\ref{eq:bern})$, we have the following expression for the terminal
speed,
\begin{eqnarray}
v_\infty ^2=&&\Bigl [{\gamma +1 \over \gamma -1} c_{s,b}^2 + 2 \pi \Gamma_0 {G M_d \over d} \Bigr ] \nonumber\\
&& -\Bigl [ {4 G M_d \over d } +{ G M_b \over a} +{2G M_s \ln (1+a /r_s) \over a} \Bigr ]\,.
\label{eq:term}
\end{eqnarray}
We emphasize that $\Gamma_0$ refers only to disk parameters 
and $\Gamma_0=1$ does not signify Eddington luminosity for 
radiation pressure on dust.  

\begin{figure}
\vspace{24pt}
\includegraphics[width=80mm]{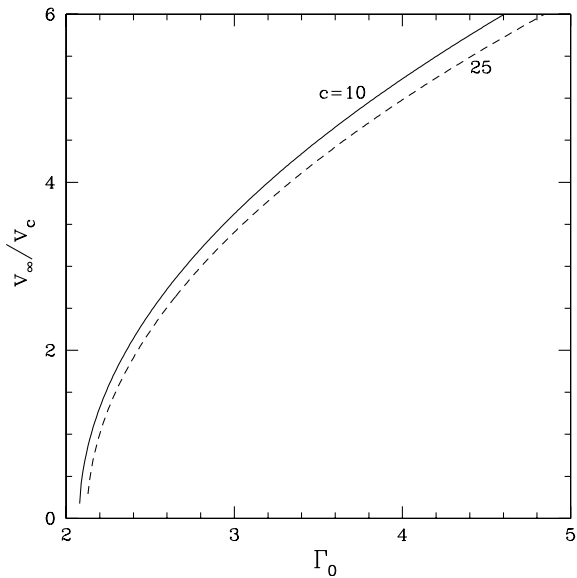}  
\caption{The ratio between $v_\infty$ and $v_c$ is shown as a function
of $\Gamma_0$ for galaxies with two different concentration parameter $c$, for cold gaseous outflow.
         }
\label{figcie}
\end{figure}

Figure 1 shows the dependence of $v_\infty/v_c$ on $\Gamma_0={\kappa I/2 \pi c G \Sigma}$, and on
the concentration parameter $c$ for $c_{s,b}=0$, and for $a=r_b$.
The sound speed  makes only a small difference to the terminal speed,
 but it helps establishing a quasi-steady regime in the
outflow redistributing the pressure over the entire flow.
The curves show that
there is a minimum value of $\Gamma_0 \sim 2$ for the wind to reach infinity.
In general the wind speeds are lower for higher $c$, which
is the case of low-mass and more compact galaxies, but the variation is small.
The important finding here is that the wind requires $\Gamma_0 > 1$. 

We next use
a relation between $c$ and galactic mass given by Macci\'o \etal (2007), and calculate the wind
speed for galaxies with different masses, or rotation speeds. One can extend the calculations
to higher redshift by using the scaling $c\propto (1+z)^{-1}$. 
We use $\Omega_0=0.27, h=0.7$, and calculate the NFW profiles for the
$\Lambda$-cosmology.
Figure 2 shows the results for
$v_\infty/v_c$ as a function of $\Gamma_0$ for different $v_c$, at $z=0$ and
at $z=7$. The curves show that the $v_\infty/v_c$
depends mostly on $\Gamma_0$ and varies weakly with rotation speed and redshift.

We next ask the question what determines the value of $\Gamma_0$ which 
is crucial for the
calculation of $v_\infty$.

\begin{figure}
\vspace{24pt}
\includegraphics[width=80mm]{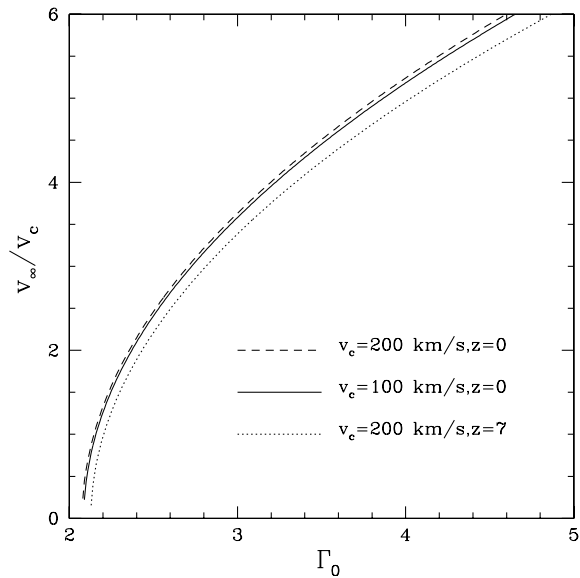}  
\caption{The ratio between $v_\infty$ and $v_c$ is shown as a function
of $\Gamma_0$ for galaxies with different $v_c$ at two different redshifts. 
         }
\label{figcie}
\end{figure}

\subsection{Evolution of wind speed with time}
We recall that the value of $\Gamma_0$ depends on disk parameters ($\Sigma, I$)
and dust grain properties (through $\kappa$). Consider first the 
value of $\Sigma/I$
that is essentially the disk mass-to-light ratio. One
can compare the observed values of mass-to-light ratio with the minimum
requirement as derived above. 
Li and Draine (2001) give the mean opacity for 
gas mixed with dust as $\sim 128$ cm$^2$ g$^{-1}$ in U and $\sim 93$ cm$^2$ g$^{-1}$ in 
the B band. Here we shall consider the B band value as a conservative estimate, which ensures that star formation is not obscured. In case of obscured star formation,
shell(s) of gas and dust would be accelerated, which are likely to undergo fragmentation
and produce clouds due to instabilities, facilitating the escape of ionizing photons
(Razoumov \& Sommer-Larsen 2010; Wise \& Cen 2009).

Using B-band values we find a maximum value of disk mass-to-light ratio required for the outflow to occur,
given by $M/L \sim \kappa/ (2cG \Gamma_0)$.
For $c\sim 10$, this is given by ${(M/M_{\odot}) \over (L_B/L_{B,\odot})}
\equiv \Upsilon_B \le 0.036$. 
This upper limit on disk $\Upsilon_B$ is much lower than observed in present day disk galaxies.
The Milky Way disk has a local value of $\Upsilon_B \sim 1.2 \pm 0.2$ (Flynn \etal 2006), 
and the typical value for disk galaxies is $1.5\pm 0.4$ (Fukugita, Hogan, Peebles 1998). 
According to Flynn \etal (2006), a third of the total disk
mass comes from gas and the rest from stars. Starburst galaxies can have a much lower value
of $\Upsilon_B$. The estimated $M/L_{bol}$ for  NGC 7714 is $\sim 0.02$ (Bernl\"ohr 1993a),
for regions in NGC 520 it is $ \sim 0.003$ (Bernl\"ohr 1993), and for a young super
cluster in M82, Smith \& Gallagher (2001) has estimated the $\Upsilon_V \sim 0.02$.
These low values of $\Upsilon$ are believed to arise from a top heavy IMF and 
young age of the stellar population (e.g., Kotilainen \etal 2001; Smith \& Gallagher 2001).

For an instantaneous burst these models predict an initial period of roughly
constant luminosity for $t \le 3$ Myr,
and a decrease in the luminosity afterwards (Bruzual
\& Charlot 2005; Buzzoni 2005; V\'asquez \& Leitherer 2005). 
 For a Salpeter IMF
and  a stellar mass-luminosity relation of the type $L \propto M^{\beta}$, the late-time
decay of the luminosity is given by $L \propto t^{-(\beta -1.35)/(\beta -1)} (\propto
t^{-0.9}$, for $\beta=3.5$). The initial
period of rather constant luminosity stems from the fact that while low-mass stars
are yet to collapse, the massive stars evolve quickly, and the duration of this
period corresponds to the main-sequence life-time of the most massive stars.

Using these models we can
determine the $\Upsilon$ for a stellar population, multiplying
 by a factor $\sim 3/2$ to
account for an additional gas mass, and determine the expected disk $\Upsilon_B$ ratio as
a function of time after an instantaneous starburst. 
Using the results in Figure 9 of V\'asquez \&
Leitherer (2005), which uses a Salpeter IMF between $0.1$ and $100$ M$_{\odot}$,
and using $M_{B,\odot}=5.45$, one can calculate $\Upsilon_B$ as a function of time.
Strength of instantaneous starburst is 
characterized by
total mass converted into stars initially, which in this case is $10^6$ M$_{\odot}$.

Using the mean dust opacity for B band, we calculated the time evolution
of $\Gamma_0$, and then determined the  wind
terminal speed using eqn$(\ref{eq:term})$. 
Although the use of eqn$(\ref{eq:term})$ assumes a constant $\Gamma_0$ along a streamline,
the distance travelled by the wind over the time-scale of change of $\Gamma_0$ ($\sim 10 Myr$)
is large ($l \ge 6.3 \, {\rm kpc} (v/600 \, {\rm km/s})$). In other words, the time-scale
for the wind to reach a considerable height above the disk is comparable to the
time-scale of evolution in $\Gamma_0$, and therefore we can use our formalism 
to estimate
the terminal speed with the evolution in $\Gamma_0$.

Figure 3 shows the evolution of $v_\infty /v_c$ with time for
 $z=0$ (solid line) and $z=7$ (dotted line).
 The curves show that the wind speed decreases rapidly after
$\sim 10$ Myr, and that $v_\infty \le 3 v_c$, its value being smaller
for compact galaxies. This result can be compared with the observed range of maximum
wind speed. Martin (2005) found that the maximum speed of clouds embedded in outflowing
gas ranges between $2\hbox{--}3 \, v_c$, and Rupke \etal (2005) found a range
of $[0.67\hbox{--}3] \, v_c$. We show this range with two dashed lines in Fig 3.
Figure 3 also shows that the wind speed is somewhat smaller at high redshift. 
The reason is
that galactic mass for a given $v_c$ is smaller at high redshift,
but $c\propto M^{-0.2} \, (1+z)^{-1}$,
and the mass
effect outweighs the redshift effect. 
However, the variation of $v_\infty/v_c$ with 
redshift is expected to
be very small in this model, much smaller than those caused by 
other parameters, such
as the IMF.

\begin{figure}
\vspace{24pt}
\includegraphics[width=80mm]{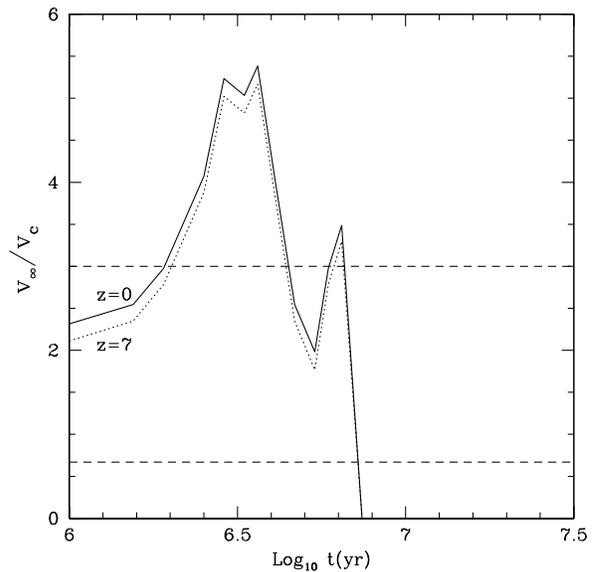}  
\caption{The evolution of terminal wind speed in terms of rotation speed with time is plotted
for $v_c=100$ km/s at $z=0$ (solid line), and at $z=7$ (dotted line). 
Horizontal dashed lines show the observed 
range at $z=0$ (Rupke \etal 2005, Martin 2005).
         }
\label{figcie}
\end{figure}

\section{Discussions}
Figure 3 shows that dust-driven winds are likely to have a terminal speed 
$\sim 2\hbox{--}3 v_c$, for a combination of reasons that involve stellar physics and the relation
between disk and  halo parameters. It is interesting that this result coincides
with observations, since there is no free parameter in our calculation. 
The strength of our approach lies in the fact that the terminal speed calculated using
the Bernoulli function is independent of the streamline used
by the gas, {\it as long as} streamlines do extend to infinity, which is our
basic assumption. 
Below we discuss a few implications.

In the scenario of energy driven winds, 
the IGM is believed to be enriched by winds from 
dwarf galaxies, since they were numerous in the early universe 
(Silk, Wyse
\& Shields 1987; Nath \& Trentham 1997; Ferrara, Pettini \& Shchekinov 2000; Cen \& Bryan 2001;
Madau, Ferrara, Rees 2001; Aguirre \etal 2001). 
However, in the case of dust-driven 
winds, the importance of low-mass galaxies in IGM enrichment diminishes because 
$v_\infty \propto v_c$. Our calculations here show that the wind speed depends strongly
on the time elapsed after a starburst, or more 
generally on the star formation history and parameters.
It is believed that the IMF is weighted towards massive
stars at high redshift (e.g., Schneider \& Omukai 2010), in which
case the wind speed
likely increases with redshift.
In this case the contribution of dwarf galaxies at high redshift may still be important.

Recent simulations for IGM enrichment including momentum driven winds 
have used the ansatz $v_\infty \sim 3 v_c$ (Oppenheimer \& Dav\'e 2006),
using the estimates of MQT05 
for wind from a spherical galaxy radiating close to the dust
Eddington luminosity.
It has been suggested that the momentum driven winds
drive a feedback loop that makes  $v_\infty \sim 3 v_c$ (Martin 2005;
MQT05). Our
calculations show that  such dust-driven winds
are possible only within a period of $\sim 10$ Myr after a starburst. 
Since the wind speed depends strongly on the value of disk $\Upsilon$, which depends on
the IMF (being smaller for a top-heavy IMF), the wind speed is expected to be larger for
a IMF skewed towards massive stars. 
The implications
needs to be studied with numerical simulations using
modified prescriptions for the wind speed. It is possible that
this puts constraints on the metallicity of the wind gas.  
In addition, as the dust opacity is proportional to the metallicity,
$\kappa\propto Z$, the existence of a threshold $\Gamma_0$ needed for the wind
to reach infinity suggests that only metal-rich galaxies can enrich
IGM.

We note that our calculation does not determine streamlines, without which
we cannot calculate the density structure in the wind, 
and therefore cannot derive the mass 
loss rate. In the case of
single scattering, the maximum mass loss rate is $\dot{M} \sim L/(v_\infty c)$
(MQT05).
This recovers the scaling that $\dot{M} \propto v_c^{-1}$ adopted by numerical simulations
(Oppenheimer \& Dav\'e 2006). Taking our results into account, the mass loss rate for
a given luminosity is expected to be lower at high redshift, because of the 
possible rise of $v_\infty$
with $z$ arising from IMF evolution.  

In summary, we have derived a terminal speed for dust-driven outflows from disk galaxies,
and have shown that $v_\infty \sim (2 \hbox{--}3) v_c$, which is determined by the minimum
value of disk $\Upsilon_B \sim 10^{-2}$ arising from the luminosity of a stellar population, and the relation
between the disk and dark matter halo that fixes the terminal speed for a given value of $\Upsilon$.
We have shown that dust-driven winds from disk galaxies are excited within a time-scale of $\sim 10$ Myr
of a starburst, after which the radiation pressure on dust is unable to drive outflows.

\bigskip

We thank Mitchel C. Begelman, J. Silk and S. Sridhar for valuable discussions.
We thank the referee C. D. Matzner for useful comments.

\end{document}